\documentclass[referee]{raa}           

\usepackage{graphicx,times}
\usepackage{natbib}
\usepackage{amssymb,amsmath}
\bibpunct{(}{)}{;}{a}{}{,}

\usepackage[pagebackref=true]{hyperref}

\begin{document}

   \title{The {\em Lobster Eye Imager for Astronomy} Onboard the SATech-01 Satellite}

 \volnopage{ {\bf 20XX} Vol.\ {\bf X} No. {\bf XX}, 000--000}
   \setcounter{page}{1}

   \author{Z.X. Ling\inst{1,2},
   	       	X.J. Sun\inst{3},
   	       	C. Zhang\inst{1,2},
   	       	S.L. Sun\inst{3},
   	       	G. Jin\inst{4},
   	       	S.N. Zhang\inst{1,2,5},
   	       	X.F. Zhang\inst{6},
   	       	J.B. Chang\inst{3},
   	       	F.S. Chen\inst{3},
		Y.F. Chen\inst{3}
% ...
   }

   \institute{ National Astronomical Observatories, Chinese Academy of Sciences, Beijing, China; {\it chzhang@nao.cas.cn}\\
        \and
       	School of Astronomy and Space Science, University of Chinese Academy of Sciences, Beijing, China\\
	\and
	Shanghai Institute of Technical Physics, Chinese Academy of Sciences, Shanghai, China;{\it sunxiaojin@mail.sitp.ac.cn}\\
	\and
	North Night Vision Technology Co., LTD, Nanjing, China\\
	\and
	Institute of High Energy Physics, Chinese Academy of Sciences, Beijing, China\\
	\and
	Science and Technology on Low-Light-Level Night Vision Laboratory, Xi'an, China\\
	\and
	Innovation Academy for Microsatellites, Chinese Academy of Sciences, Shanghai, China\\
	\and
	Institute of Electrical Engineering, Chinese Academy of Science, Beijing, China\\
	\and
	National Space Science Center, Chinese Academy of Science, Beijing, China\\
\vs \no
   {\small Received 20XX Month Day; accepted 20XX Month Day}
}

\abstract{The {\em Lobster Eye Imager for Astronomy} ({\em LEIA}), a pathfinder of the Wide-field X-ray Telescope of the Einstein Probe (EP) mission, was successfully launched onboard the SATech-01 satellite of the Chinese Academy of Sciences on 27 July 2022. In this paper, we introduce the design and on-ground test results of the {\em LEIA} instrument. Using state-of-the-art Micro-Pore Optics (MPO), a wide field-of-view (FoV) of 346 square degrees (18.6 degrees $\times$ 18.6 degrees) of the X-ray imager is realized. An optical assembly composed of 36 MPO chips is used to focus incident X-ray photons, and four large-format complementary metal-oxide
semiconductor (CMOS) sensors, each of 6 cm $\times$ 6 cm,  are used as the focal plane detectors. 
The instrument has an angular resolution of 4 -- 8 arcmin (in FWHM) for the central focal spot of the point spread function, and an effective area of $2-3$ cm$^2$ at 1 keV in essentially all the directions within the field of view. 
The detection passband is 0.5 -- 4 keV in the soft X-rays and the sensitivity is $2 - 3 \times 10^{-11} \rm{\ erg\ s^{-1}\ cm^{-2}}$ (about 1 mini-Crab) at 1,000 second observation.
The total weight of {\em LEIA} is 56 kg and the power is 85 W. 
The satellite, with a design lifetime of 2 years, operates in a Sun-synchronous orbit of 500 km with an orbital period of 95 minutes.
{\em LEIA} is paving the way for future missions by verifying in flight  
the technologies of both novel focusing imaging optics and CMOS sensors for X-ray observation, and by optimizing the working setups of the instrumental parameters.
In addition, {\em LEIA} is able to carry out  scientific observations to find new transients and to monitor known sources in the soft X-ray band, albeit limited useful observing time available.
}

   \authorrunning{Zhixing Ling et al. }            
   \titlerunning{The {\em Lobster Eye Imager for Astronomy}}  
   \maketitle

%
\section{Introduction}
\label{sec:introduction}

X-ray wide-field monitoring observations in time-domain have been playing an important role in discovering and understanding the X-ray Universe. 
The Wolter-I telescopes, which are commonly used in X-ray astronomy, are inconvenient for use as monitoring-type instruments, however, due to their small field-of-view (FoV) of typically 1 degree or less.
An interesting lobster-eye design was proposed by 
\cite{1979ApJ...233..364A}, which can in principle realize both a large FoV and relatively high sensitivity, by making use of micro-pore optics (MPO). 
In theory, the imaging properties of such an optic remain invariant in essentially all the directions across the entire FoV except at close to the edges. This means that, theoretically, there is almost no vignetting effect over most of the FoV. 
In the past decades, extensive efforts have been made in the study of lobster-eye optics and the development of the MPO techniques \citep{1989RScI...60.1026W,1991RScI...62.1542C,1992SPIE.1546...41F,1992ApOpt..31.7339K,1993NIMPA.324..404F,1996ApOpt..35.4420P,1999NIMPA.431..356B,2015RScI...86g1301C,2016SPIE.9905E..1YW,2017SPIE10567E..19H}.
Despite of these efforts, no in-flight observations made by X-ray focusing lobster-eye telescopes with a truly large FoV (i.e. the order of 100 square degrees or larger) have been achieved until recently.

The Einstein Probe (EP) \citep{2016SSRv..202..235Y,2018SPIE10699E..25Y,Yuan2022}
of the Chinese Academy of Sciences (CAS) is a space mission for time-domain astronomy equipped with a wide-field X-ray monitor built from lobster-eye MPO.
The EP mission is an international collaborating project led by the CAS with participation from the European Space Agency (ESA), the Max-Planck Institute for Extraterrestrial Physics (MPE), and the France Space Agency (CNES).
The mission was formally adopted in December 2017 and is aimed for launch at the end of 2023. 
EP has two payloads, a wide-field X-ray telescope (WXT) making use of lobster-eye MPO (Ling et al. in prep.) and a follow-up X-ray telescope (FXT) of the Wolter-I type \citep{2020SPIE11444E..5BC}.
EP-WXT is composed of 12 identical modules and has an overall FoV as large as 3600 sq.deg. 
Besides lobster-eye MPO, anther novel technique employed for WXT is a large array of CMOS sensors  used as focal plane detectors\citep{Yuan2022}

A pathfinder for the EP-WXT and telescopes of this kind was proposed in 2019 by the EP team.
The aims are to verify the in-orbit performance of both the MPO and CMOS components used for the EP-WXT, and to optimize the operational instrumental parameters in space.
The pathfinder was successfully launched on 2022 July 27, piggybacked as one of the scientific payloads onboard the SATech-01 satellite.
The instrument was named as Lobster Eye Imager for Astronomy ({\em LEIA}) afterwards.
{\em LEIA} will soon complete the performance verification and in-orbit calibration phase and is starting to carry out scientific observations as a wide-field X-ray monitor.
This paper introduces the {\em LEIA} instrument. The design of {\em LEIA} is presented in Section 2; the simulated performance and the ground test result are summarized in  Section 3. The {\em LEIA} experiment is described in Section 4. The result of {\em LEIA} and current status are shown in Section 5, followed by a summary in Section 6.

\section{Instrumental design}
\label{basic_property}

\begin{figure}[htbp]
\centering
\includegraphics[width=0.6\textwidth]{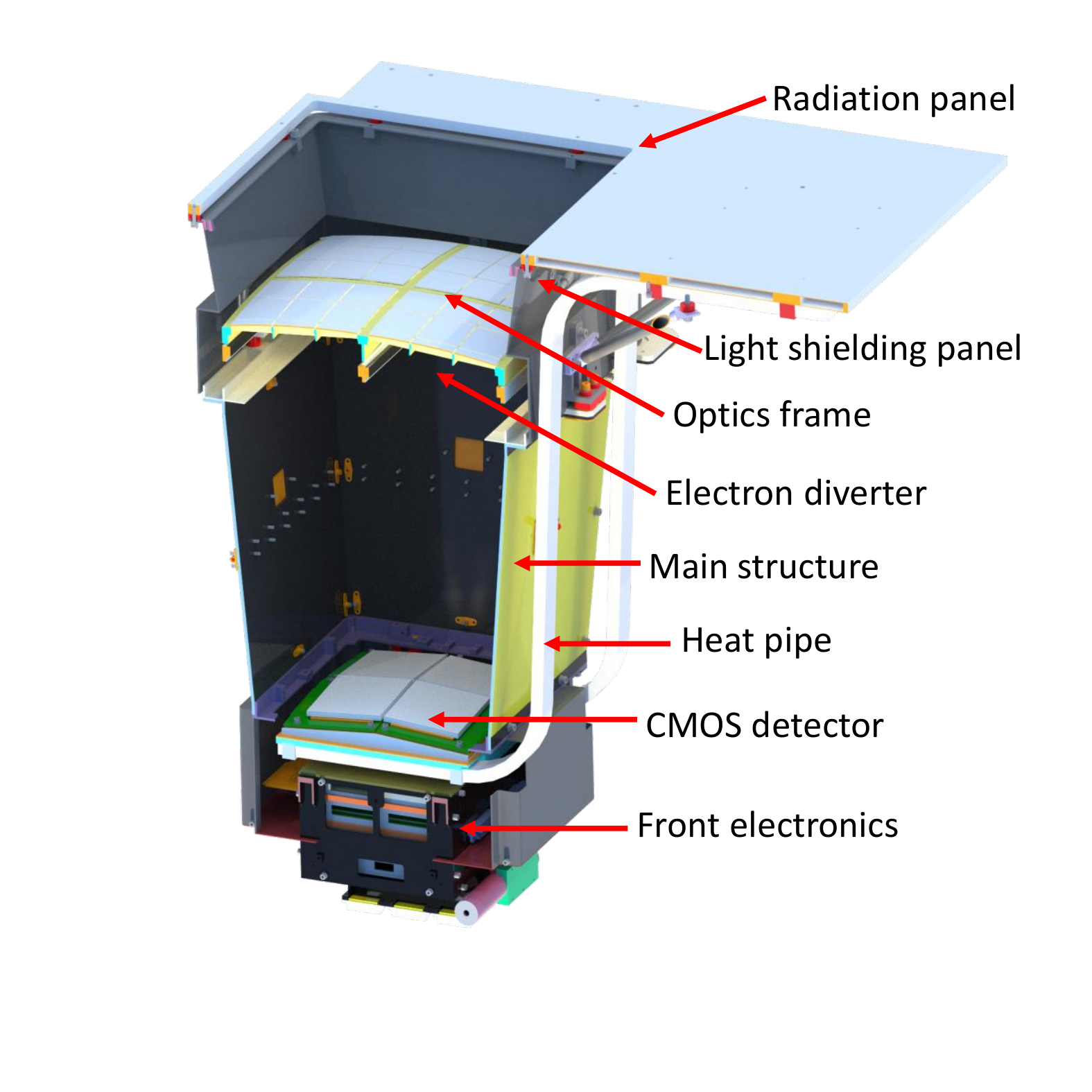}
\caption{Schematic view of {\em LEIA}.}
\label{LEIA_figure}
\end{figure}

{\em LEIA} is a test model of one of the 12 WXT modules onboard the Einstein Probe satellite.
Here we summarize the basic design of {\em LEIA}. For an account of the technical details of the WXT instrument, see Ling et al. (in preparation).  
To achieve both a large FoV and focusing imaging, {\em LEIA} uses Lobster-eye MPO to focus incident X-ray photons. 
An array of mosaics of 6 $\times$ 6 = 36 plates, each of 4.25 cm $\times$ 4.25 cm in size and 2.5 mm thickness, forms the focusing optic. The designed focal length of the MPO optics is 375 mm, with a deviation of about $\pm 1.5\%$.
The optic system is divided into four quadrants, each composed of 3 $\times$ 3 MPO plates. On the focal sphere corresponding to each of the optic quadrants, a large-format CMOS sensor of 6 cm $\times$ 6 cm in size is used as a detector to collect X-ray photons. Although the CMOS detectors are planer, they are tilted with respect to each other on the focal plane to best fit the curved focal surface. Figure \ref{LEIA_figure}. shows the shape of the focal plane. The total FoV of the imager is 18.6 degrees $\times$ 18.6 degrees. A schematic view of the imaging system can be found in Figure 1 (left panel) in \cite{Zhang2022}.
 
  To our knowledge, there had been no CMOS sensors operated as X-ray astronomical detectors in orbit prior to {\em LEIA}. 
  The CMOS sensors are back-illuminated, with a read-out frame rate of 20 Hz. 
 The pixel size is 15 $\mu$m $\times$ 15 $\mu$m, much smaller than the angular resolution of the mirror assembly. 
 The X-ray performance of the CMOS sensors very similar to those used for {\em LEIA} (the only difference is that the ones used for {\em LEIA} are coated with a layer of aluminum) can be found in \cite{2022PASP..134c5006W}.
 The built-in thermal control of the module is applied to keep the temperature of the optic assembly at $\rm{6  ^{\circ}\!C}$. 
 The CMOS detectors are designed to operate at a temperature of $\rm{-30 ^{\circ}\!C}$, which is optimised to achieve good detector performance while minimising the demand for power resource.
 To prevent optical and ultraviolet light from reaching the detectors in orbit, a 200 nm-thick aluminum layer is coated on the surface of the CMOS sensors.
 Moreover, an 80 nm-thick aluminum layer is also coated directly on the MPO plates. 
 The specifications of {\em LEIA} are listed in Table \ref{table_LEIA}.

\begin{table}[h!]
\centering
\begin{tabular}{ l | l  }
\hline
Optic & 36 MPO plates \\ \hline
Detector & Four 6 cm $\times$ 6 cm CMOS sensors \\ \hline
Field of View & 346 square degrees  \\ \hline
Focal length & 375 mm \\ \hline
Angular resolution & 4 -- 8 arcmin (FWHM, measured) \\ \hline
Energy band & 0.5 -- 4 keV  \\ \hline
Energy resolution &  125 $\pm$ 2 eV at 1.25 keV  \\ \hline
Time resolution & 50 ms  \\ \hline
Source positioning accuracy$^{\rm a)}$ & \textless \ 1 arcmin \\ \hline
Weight  &  53 kg (total) \\ \hline
Power &  85 W \\
\hline
\end{tabular}
\caption{Instrument specifications of {\em LEIA}.}
\label{table_LEIA}
Notes: a) In the detector coordinates.
\end{table}

The design of the imager is shown in Figure \ref{LEIA_figure}. 

\begin{figure}[htbp]
\centering
\includegraphics[width=0.4\textwidth]{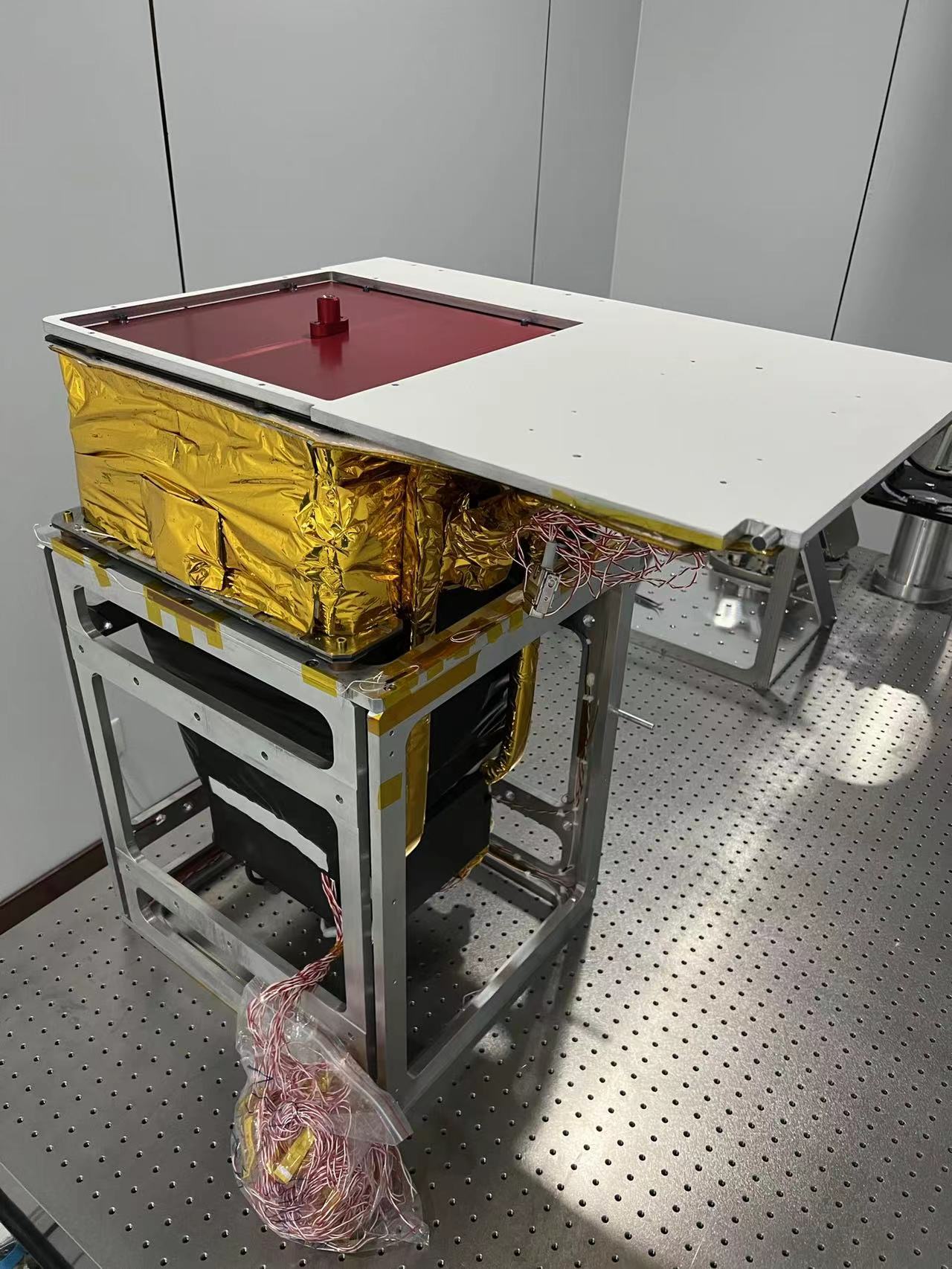}
\caption{Picture of the {\em LEIA} instrument after engineering integration.}
\label{LEIA_photo}
\end{figure}

At the top is a radiating panel that is connected to the detector unit with a heat pipe to cool the detectors down to the required working temperature (-30 degrees). An optical baffle is mounted above the optic assembly, connecting the radiation panel. An electron diverter is mounted beneath the optic assembly to deflect low-energy electrons in orbit. 
The deflecting efficiency is greater than 97$\%$ for electrons of energy less than 1 MeV \citep{wanglei2020}.
 The main mechanical structure connects the optics assembly and the focal plane unit. The detector and its front-end electronics box are located at the bottom of the instrument. The vibration frequency of the whole structure is 280 Hz, far beyond the first-order frequency of the satellite platform. This main component of the imager weighs 26 kg with a power of 25 W. The overall size is 695 mm $\times$ 540 mm $\times$ 698 mm.
An exterior view of the lobster-eye imager is shown in Figure \ref{LEIA_photo}. 

In addition to the upper main structure, an independent and auxiliary electronics control box (ECB) is attached to the module inside the satellite platform for the control of the instrument and its temperature. A transient triggering board is integrated into the ECB to realize the algorithm to find new transients in real time. 
The alert message of triggers will be transmitted to the ground by the Beidou navigation system in real time.  
The ECB will package all the scientific and housekeeping data and send them to the satellite storage disk. The estimated data rate for {\em LEIA} is about 5 GB per day. The weight of the ECB is 27 kg and the power is 60 W, making a total weight of 53 kg and a total power of 85 W for the whole payload. 

{\em LEIA} was developed jointly by the National Astronomical Observatories (NAOC) and the Shanghai Institute of Technical Physics (SITP), CAS. 
The mirror assembly was developed at the X-ray Imaging Lab of NAOC, and the complete WXT module was engineered by integrating the mirror assembly, the detector unit, and thermal control at SITP.
The MPO plates were manufactured by North Night Vision Technology Company (NVVT) in collaboration with and jointly sponsored by the CAS and the National Natural Science Foundation of China.

\section{Performance and on-ground calibration}

\subsection{Simulated performance}

\begin{figure}[htbp]
\centering
\includegraphics[width=1.0\textwidth]{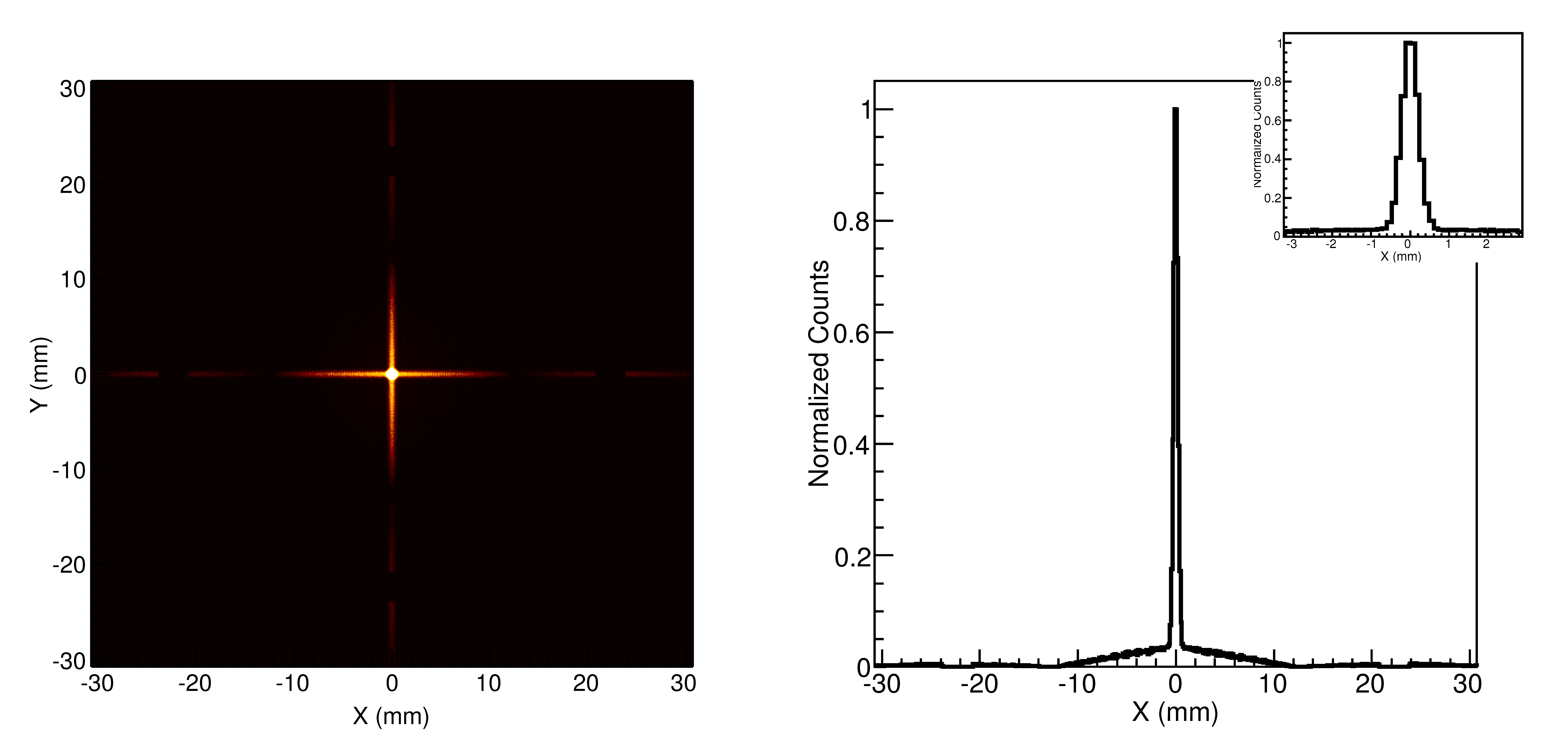}
\caption{The simulated PSF at 1 keV for the {\em LEIA} instrument. The focus length is 375 mm of the MPO optical assembly.}
\label{psf}
\end{figure}

\begin{figure}[htbp]
\centering
\includegraphics[width=0.8\textwidth]{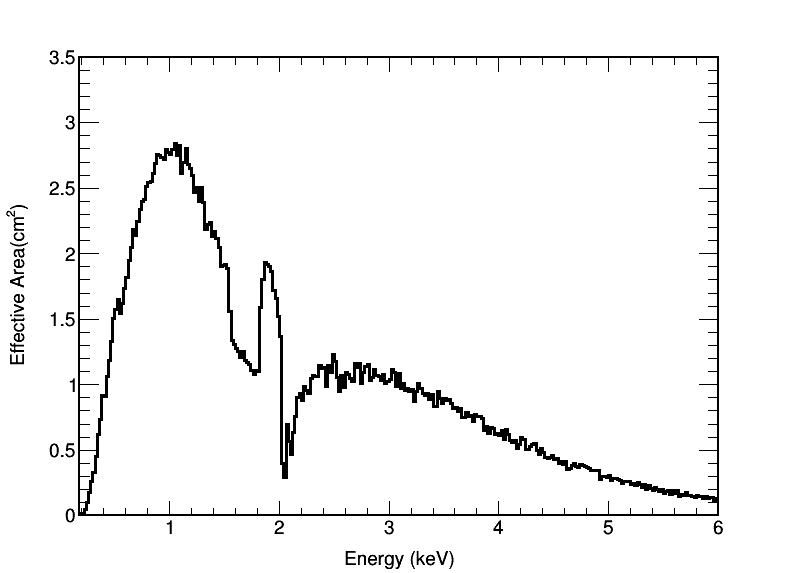}
\caption{The total effective area of {\em LEIA} instrument with a central spot of radius of 6 arcmin. This area including the mirror assembly and the quantum efficiency of detector. The dip near the 1.5 keV is due to the absorption of the aluminum coating. The increase near the 1.8 keV is due to the absorption edge of silicon. The dip near 2 keV is due to the M absorption edge of Iridium. A thin layer of  Iridium is coated on the channel surface of the MPO chips to boost the reflection efficiency of soft X-ray.}
\label{effarea}
\end{figure}

The performance of the EP WXT module has been studied extensively by both simulations and on-ground tests and calibration.
A Geant4 model was built to simulate its performance and the background in orbit \citep{2017ExA....43..267Z}.
The simulated image of the point spread function (PSF) and the effective area of the optics at 1 keV are shown in Figure \ref{psf} and Figure \ref{effarea}, respectively. 
As a distinctive feature of the lobster-eye MPO optics, the PSF is of a cruciform shape with a bright central focal spot \citep{1979ApJ...233..364A}.
The size of the central focal spot is commonly used to represent the PSF and hence the angular resolution of the lobster-eye optics. 
 The angular resolution of the mirror assembly is expected to be around 5 arcmin in full width at half maximum (FWHM) in most directions within the FoV. The sigma of a Gaussian fitting of the core region is about 1 to 2 arcmin, as shown in the right bottom panel of Figure \ref{psfmeasure}. 
 This PSF value leads to a source localization accuracy of 1 arcmin for a source with about 200 counts (statistical errors in the detector coordinates).  
Using a photometric aperture of a circle of 6 arcmin radius (corresponding 3 $\sigma$ of the core region of PSF) around the central spot, the simulated effective area is about 3.5 $\rm{cm^2}$ at 1 keV (for the optics only, without taking into account the focal plane detector).
This is consistent with the on-ground calibration result (Cheng et al. in prep.). Figure \ref{effarea} shows the effective area of {\em LEIA} in the energy band 0.3 - 6 keV. 

The in-orbit background, including the cosmic diffuse X-ray emission and detector background, has also been simulated by considering the charged particle environment. It should be noted that in these simulations, the orbit of the EP satellite was adopted, which is different from that of {\em LEIA}. The diffuse cosmic X-ray photons contribute 0.14 $\rm{count\ s^{-1}\ cm^{-2}}$ to the background, whereas the particle-induced background is 0.17 $\rm{count\ s^{-1}\ cm^{-2}}$ , resulting in a total background rate of $\sim$ 0.31 
$\rm{count\ s^{-1}\ cm^{-2}}$ . Based on these background results, the expected sensitivity of {\em LEIA} is estimated to be about 2 - 3 $\times$ $ 10^{-11} $$\rm{\ erg\ s^{-1}\ cm^{-2}}$ , or $\sim$ 1 mini-Crab for a typical observation duration of 1000 seconds\footnote{http://ep.bao.ac.cn}.
The details and results of the background simulation can be found in \cite{2018SPIE10699E..5NZ}.

The oversampling of the point spread function with the small pixel size, the high frame rate of the CMOS detectors, and the small effective area make the instrument almost free from pile-up effects even for X-ray sources as bright as Sco X-1. 

\subsection{Instrument tests and calibrations} 

The instrument tests and calibrations were carried out in two stages. Firstly, the lobster-eye mirror assembly and the CMOS sensors were tested and calibrated separately. When the complete telescope module was engineered, end to end tests and calibrations were performed in order to achieve a thorough understanding of the instrument performance. 
The detailed setup, procedure and results of these calibration experiments are topics of separate papers and will be presented elsewhere. 
In this paper, we briefly introduce the tests and calibrations carried out for {\em LEIA} and summarize the main results only. 

\begin{figure}[htbp]
\centering
\includegraphics[width=1 \textwidth]{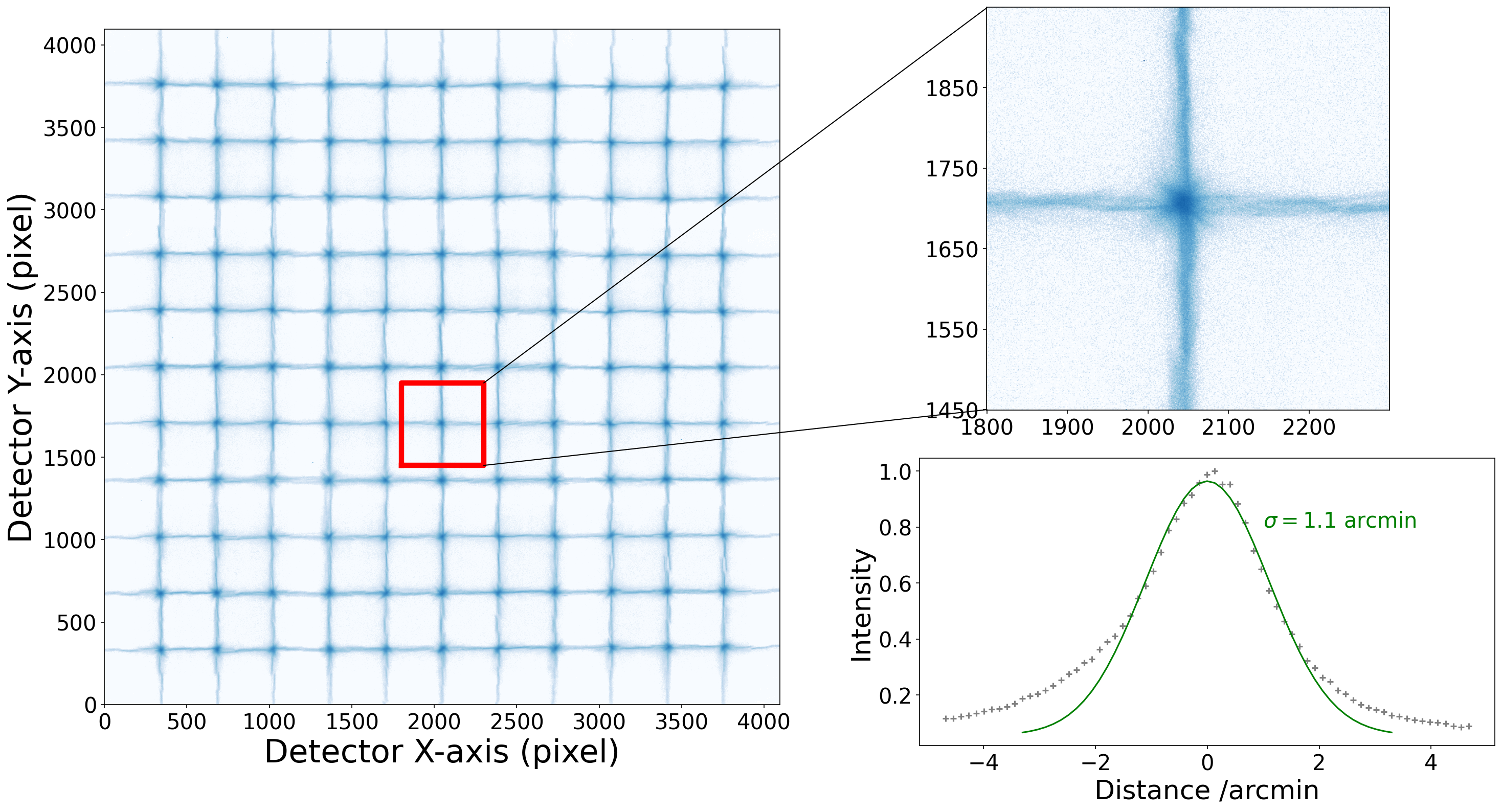}
\caption{Mosaicked image of an 11 $\times$ 11 array of the images of the PSF of the first quarter of the optical assembly at NAOC. The right top panel shows the zoom-in image of a typical PSF. The right bottom panel shows the 1D profile  of the PSF. Here, we use a Gaussian function to fit the central part of the profile. The fitted gaussian standard deviation is only 1.1 arcmin. }
\label{psfmeasure}
\end{figure}

The mirror assembly was calibrated at the PANTER facility of MPE
\citep{2005ExA....20..405F}
in April 2021 by a joint team from MPE and University of Leicester under the collaboration framework with ESA. 
Afterwards, it's optical performance was tested twice at NAOC, one before and another after a qualification vibration test in September 2021. 
In these two tests, the mirror assembly was illuminated by an X-ray beam from a point-like X-ray source about 15 meters away. A Ti target X-ray tube is used to produce a soft continuum X-ray spectrum with a high voltage setting of 6 KV.
The source is scanned in an 11 $\times$ 11 array of different directions within the FoV of each of the four quadrants by changing the aspects of the mirror assembly.  
Figure \ref{psfmeasure} shows a mosaicked 11 $\times$ 11 array of the images of the PSF over one quadrant of the FOV of the mirror assembly. The right top panel of Figure \ref{psfmeasure} shows a typical PSF of the optics. The right bottom panel shows the projection of the PSF. A Gaussian function to fit the central part of the projection to represent the angular resolution (The king function could fit the curve better, especially in the shoulder of the curve). 
The measured PSF values are in the range of 4 -- 8 arcmin in FWHM and about 85\%  less than 5 arcmin. 
The measured PSF are generally consistent with the Geant4 simulation results.

\begin{figure}[htbp]
\centering
\includegraphics[width=0.8\textwidth]{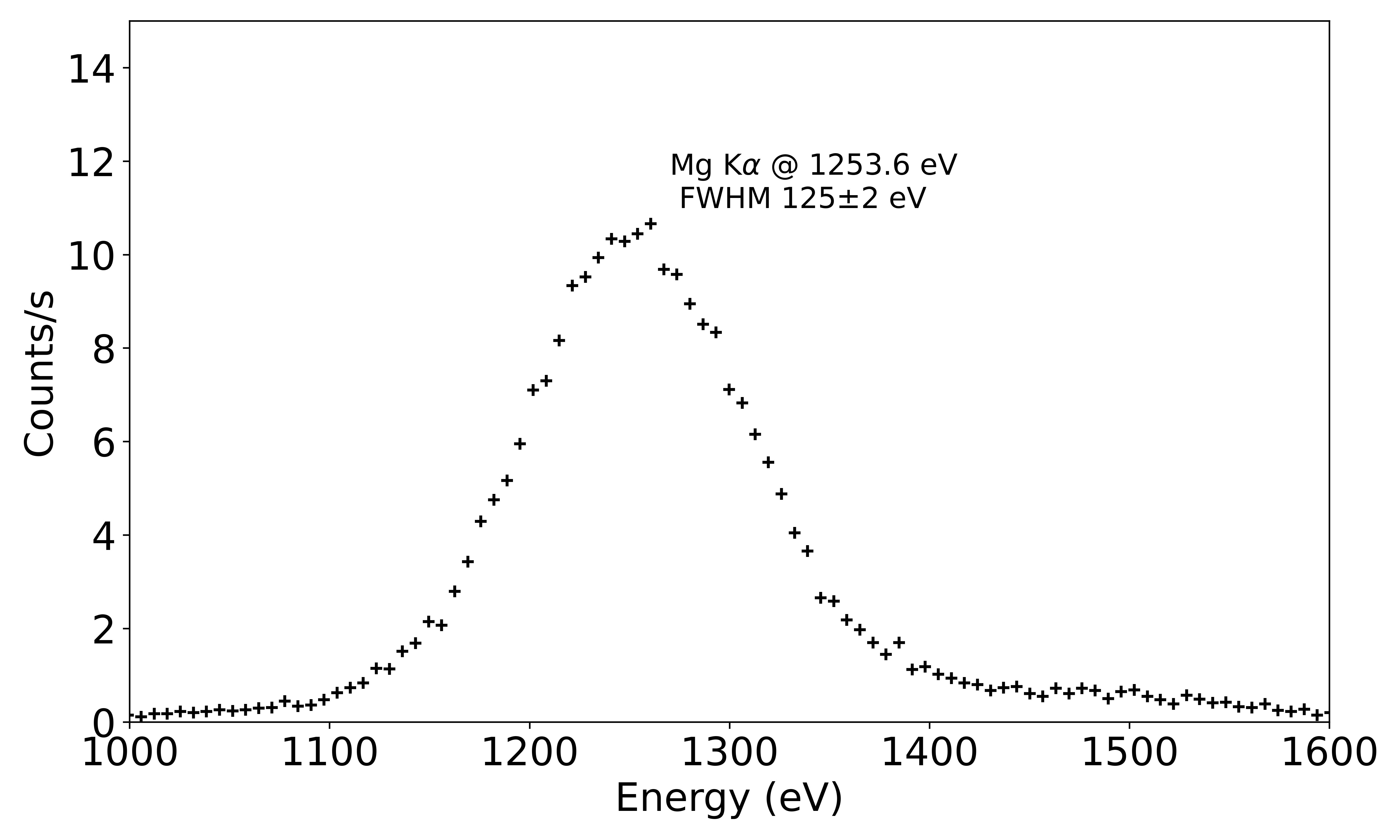}
\caption{Spectrum of the Mg Ka line at 1.25 keV measured in the calibration undertaken at IHEP/CAS.
The energy resolution is 125 $\pm$ 2 eV.}
\label{Mgkaline}
\end{figure}

Unfortunately, for three of the overall 36 MPO plates, their foci were found to be misaligned from the focal point of the module, due to the incorrect plate parameters used in the mounting. 
However, the effect of these misalignments is limited and affects the imaging quality of two localized areas in two quadrants, which amounts only 10\% of the total FoV of the module. This misalignment could be corrected by a series of full scan observations across the whole FoV.

In 2021 December, an end-to-end test and calibration campaign for the complete module was carried out at the 100m test facility 
\citep{2022ExA...tmp...76W}
at Institute of High-energy Physics (IHEP), CAS. 
The campaign includes calibration of the imaging quality and PSF, imaging offset, effective area, energy resolution.
X-ray photons from emission lines of several targets, including Mg, Cu, Si, and Ti, were applied. 
X-ray events were extracted from the CMOS sensors, which show various patterns of pixels hit by photons (see \cite{2022PASP..134c5006W} for the definition of the event pixel patterns for CMOS sensors).  
Single, double, triple, and quartic events are recorded as true X-ray events. Other grades are excluded during data extraction. The energy resolution (FWHM) is 125 $\pm$ 2 eV at 1.25 keV and 175 $\pm$ 2 eV at 4.5 keV. 
As an example, the detector spectrum of the Mg Ka line (1.25 keV) is shown in Figure \ref{Mgkaline}. 
The transmission of optical light through the aluminum filters, for overall the layers on the CMOS and MPO, was measured to be less than $10^{-6}$.

The effective area of the complete module was also calibrated in a 11 $\times$ 11 grid of directions.
The measured effective areas are in the range of 2 -- 3 $\rm{cm^2}$ at 1.25 keV across the entire FoV, except at the edges of FoV.
Theoretically, ideal lobster-eye optics predict the same PSF shape and effective area in all the directions (except at the edge).   
The mild non-uniformity, i.e. a decrease of the effective area or a change of the PSF in some of the directions, across the FoV, as measured for this instrument, can easily be understood and can be reproduced by simulations. The non-uniformity of the PSF arise from these factors: the imperfectness during the manufacturing and the mounting of the optics, the mismatch between the flat detector plane and the spherical focal plane. The obscuration of the mounting frame between the individual MPO plates for the incident X-ray photons would produce a smaller effective area in some directions than the
nominal $\sim$3 $\rm{cm^2}$ of the MPO plates, causing the non-uniformity of effective area except for the edge region.

To summarize, results of the on-ground calibrations are generally consistent with those of the simulations. 
The distributions of the PSF and effective areas are basically uniform (showing no substantial variations) across the whole FoV, as expected for a lobster-eye optic. The deviations from the prediction of an ideal lobster-eye telescope is mostly caused by imperfections inherent in manufacturing process of the MPO plates and the mounting, and the obscuration of the MPO supporting structures of the mirror assembly.

\section{The LEIA experiment}

\begin{figure}[htbp]
\centering
\includegraphics[width=0.8\textwidth]{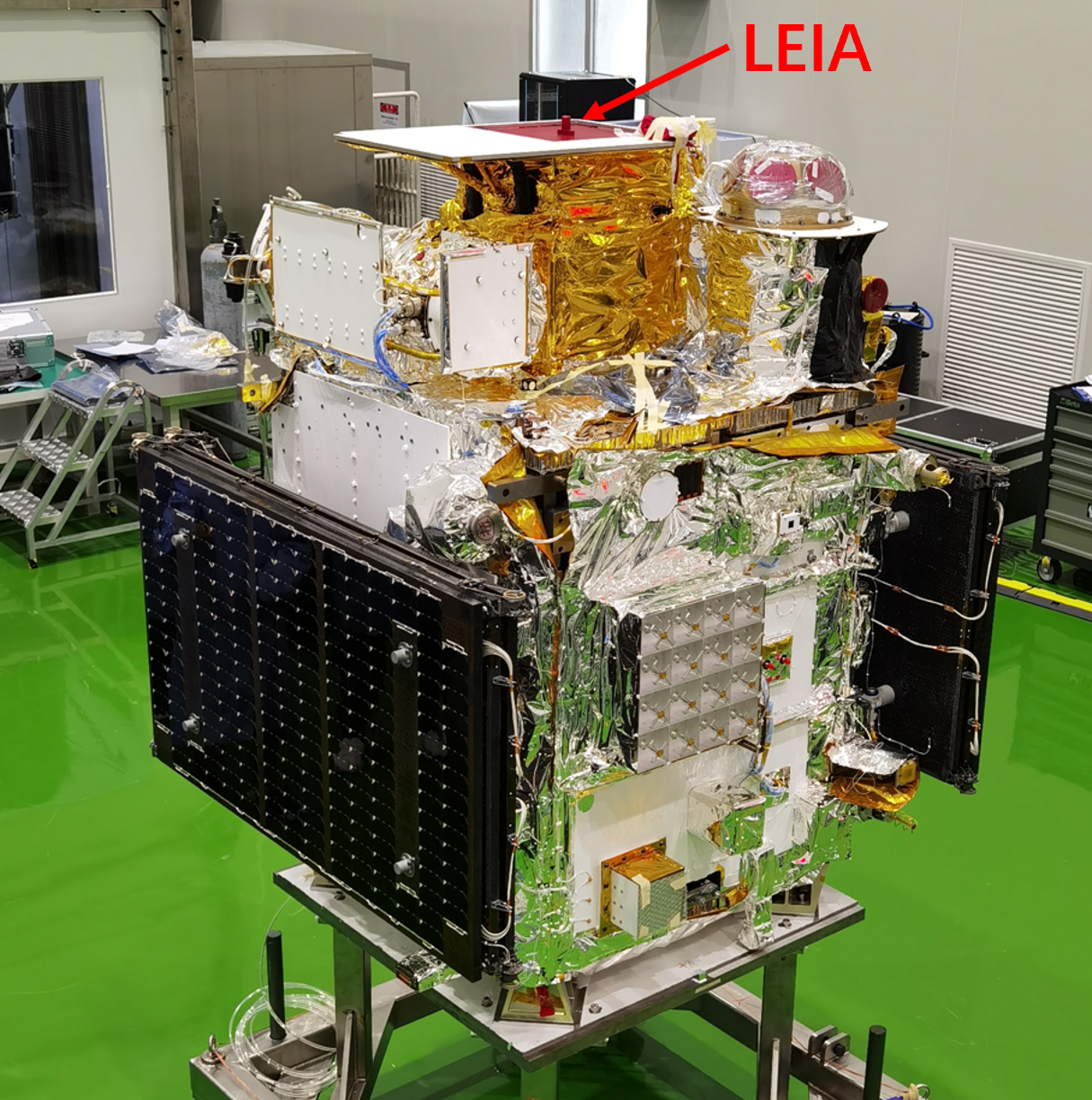}
\caption{The photograph of the SATech-01 satellite.}
\label{SATech01}
\end{figure}

The {\em LEIA} instrument was successfully launched on 2022 July 27 carried by a solid-fueled "LiJian" rocket from Jiuquan satellite launch center in China, 
piggybacked onboard the SATech-01 satellite.
With a total of 16 scientific instruments onboard, SATech-01 is
developed by the Innovation Academy for Microsatellites of CAS.
Figure \ref{SATech01} shows a picture of SATech-01 before launch, in which {\em LEIA} is sitting on top of the satellite.
In addition to {\em LEIA}, there are other two instruments for astronomical research, the {\em Solar Upper Transition Region Imager (SUTRI)} (Bai et al. in prep.) for solar observations in optical 
and {\em High Energy Burst Searcher (HEBS)} to detect gamma-ray bursts (Xiong et al. in prep.).
The parameters of SATech-01 are listed in Table \ref{table_SATech-01}. 
The satellite has a total weight of 620 kg, a total power of 480 W, and a storage capacity of 2 T bit. 
The satellite is in a Sun-synchronous orbit of 500 km height, with a 95-minute orbital period. The descending node of the orbit is 10:30 AM. 
The pointing accuracy is 0.1 degrees and the stability is 1.8 arcsec per second during observation. The design life of the satellite is 2 years.

The {\em LEIA} was degassed in the first a few days after launch. 
In the first two months of the satellite's test stage, all of the 16 scientific instruments were powered on in turns. 
{\em LEIA} had a test period of 4 days at this stage.
 After about two months of the commissioning and experiments of the satellite platform and the payloads, the majority the instruments completed their tasks and became inactive, whilst the three astronomical instruments continue to operate for further testing as well as long-term scientific observations.

\begin{table}[h!]
\centering
\begin{tabular}{ l | l }
\hline
Orbit & 500 km  \\ \hline
Descending node & 10:30 A.M. \\ \hline
Total weight & 620 kg \\ \hline
Power & 480 W \\ \hline
Pointing accuracy & 0.1 degree \\ \hline
Star tracker accuracy & 3 arcsec (3 $\sigma$) \\ \hline
Stability & 1.8 arcsec/s  \\ \hline
Storage capacity & 2 T bits \\ \hline
Number of instrument & 16 \\ \hline
Dimension  & $\phi$ 963~mm (Y) $\times$ 966~mm (Z) $\times$ 1306~mm (X) \\ \hline
Life time & 2 years \\ \hline
Launch date & July 27 2022 \\ \hline
%

\end{tabular}
\caption{Basic specifications of the SATech-01 satellite.}
\label{table_SATech-01}
\end{table}

\begin{figure}[htbp]
\centering
\includegraphics[width=1.0\textwidth]{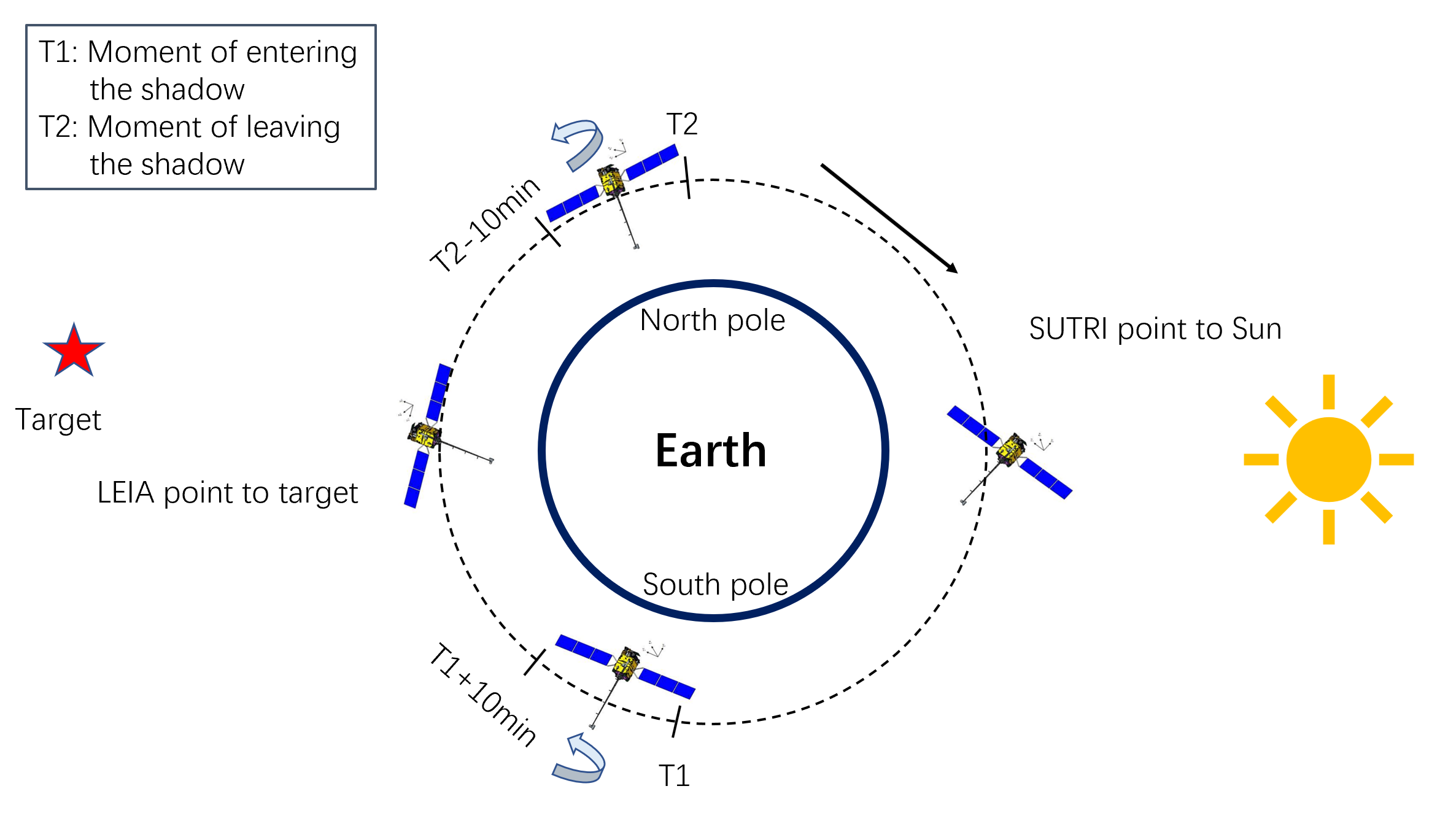}
\caption{The operation mode of the SATech-01 satellite.}
\label{SATech01orbit}
\end{figure}

To protect the instrument and minimize the impact of the radiation from the Sun, a 
Sun-avoidance angle of 90 degrees is set for {\em LEIA}.
{\em LEIA} observations are carried out during the orbital phase in the Earth's shadow, whilst on the day side the SUTRI instrument (whose FoV is always perpendicular to that of {\em LEIA}) is operating and pointed to the Sun.
Figure \ref{SATech01orbit} shows a schematic sketch of the satellite and {\em LEIA} operation in one orbit.
To prevent the Sun from entering the {\em LEIA} FoV during satellite slew, the satellite starts slewing 10 minutes after entering the Earth's shadow region and 10 minutes before leaving it. The maximum slew time of the spacecraft is 10 minutes. 
Only pointed observation mode is adopted so far, and usually only one patch of the sky is observed  in one orbit. 

The instrument stops to collect data in regions where extraordinarily high detector background is produced by charged particles in orbit, mainly the South Atlantic Anomaly (SAA) region and near the south and north polar regions of the Earth.
Moreover, in-orbit data show that, within considerable intervals of one orbit the particle background dominates the detected events of the detector, and have to be discarded.
Only data taken in the low geomagnetic latitude regions are not subject to high particle background. 
When all of those factors are taken into account, the useful observation time usually ranges from 10 to about 15 minutes per orbit.
The data of backgrounds induced by charged particles in {\em LEIA} observations are being analyzed and will be presented elsewhere. 
There are about 10 orbits per day available for {\em LEIA} observations. 

\begin{figure}[htbp]
\centering
\includegraphics[width=1 \textwidth]{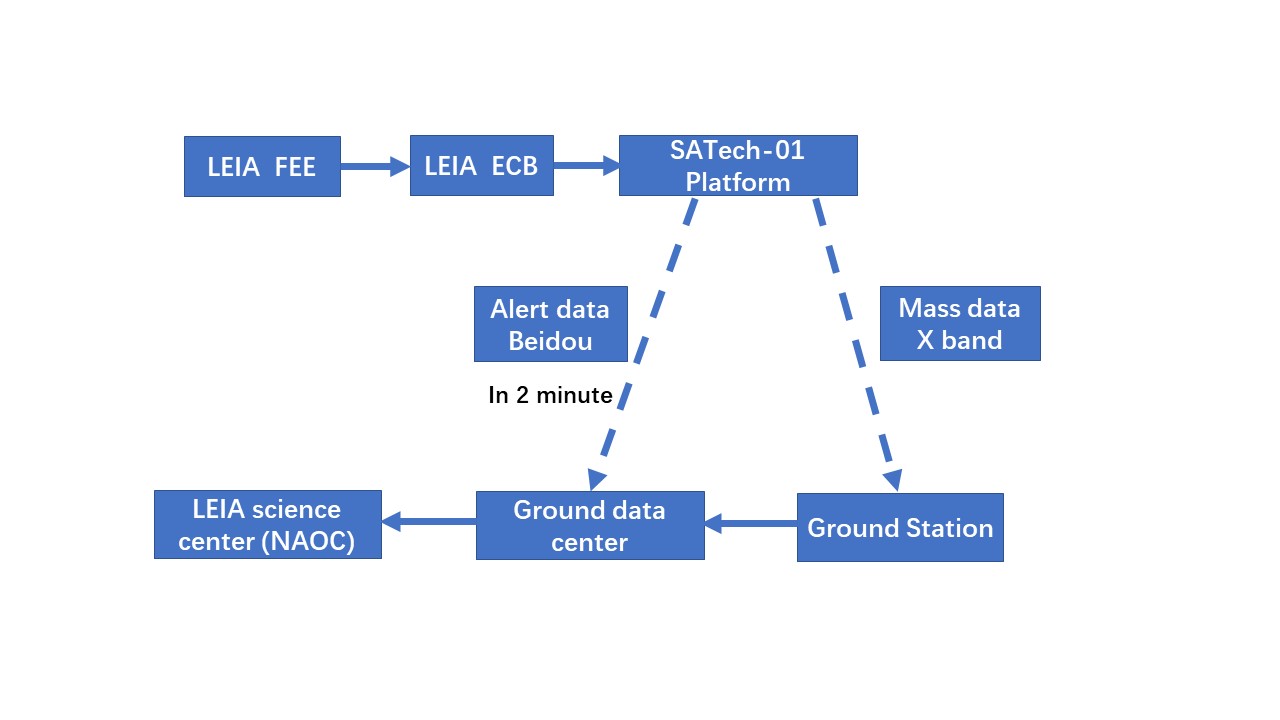}
\caption{The data flow chart of the LEIA instrument.}
\label{flowchart}
\end{figure}

There are usually four telemetry orbits per day for the command uplinking and two orbits for mass data downloading, each with a link time of about 10 minutes. The observation schedule commands are uploaded during these four orbits. In normal operation, the observation schedule is formulated once per week, and the corresponding command is uploaded every day. An urgent observation requirement could be approved in less than 2 hours and could be uploaded in the next control orbit as soon as possible. The general data rate for {\em LEIA} is about 5 GB per day. 
Figure \ref{flowchart} shows the data flow chart of {\em LEIA} instrument. The raw data will be sent to the ground station by X-band antenna during the command orbit. The trigger message could be sent to the ground data center immediately, the maximum delay is less than 2 minutes.

{\em LEIA} produces two types of scientific data. The event data record the X-ray events detected and are used for science research. The raw images of the detector can also be recorded and downloaded to periodically examine the status of the CMOS sensors and the performance of the optical blocking layers. 
These two types of data could not be recorded simultaneously.

\section{Initial results, in-orbit calibration and current status}

\subsection{Initial results}
In the first two months of operation in orbit, the instrumental performance verification (PV) and calibration were performed. 
In this section, we present selected initial results in more detail obtained in the PV phase, in addition to the first-light images released in \cite{Zhang2022}.
As the first lobster-eye wide-field focusing X-ray telescope, {\em LEIA} tested its performance by taking images of a number of celestial objects. 
The targets observed in the early PV phase include the Galactic center where the number density of bright X-ray sources is the highest, Sco X-1, Cas A, the Cygnus Loop, and the Crab nebula.
The images of the first-light observations, including part of the Galactic center region, Sco X-1 and the Cygnus Loop nebula, have been presented in \cite{Zhang2022}.
To further demonstrate the unique wide-field imaging capability of lobster-eye MPO optics, we show in Figure \ref{gc7} the X-ray image of part of the Galactic center region, in $28\deg \times 28 \deg$.
The image is mosaicked and stacked from data taken in 7 {\em LEIA} observations with a total exposure of 4652 seconds, with part of the FoVs overlapping each those of the others.
In just several observations of about 10 minutes each, the Galactic bulge region spanning $28\deg \times 28 \deg$ can be covered, with about 11 X-ray sources detected. 
The image shows no significant variations of the imaging quality and the effective area across the FoV, as predicted based on the results of the on-ground calibration. However, quantitative comparisons with the on-ground calibration have to await results of extensive in-orbit calibration, for which the observations have been performed and are still to be completed.

\begin{figure}[htbp]
\centering
\includegraphics[width=0.6 \textwidth]{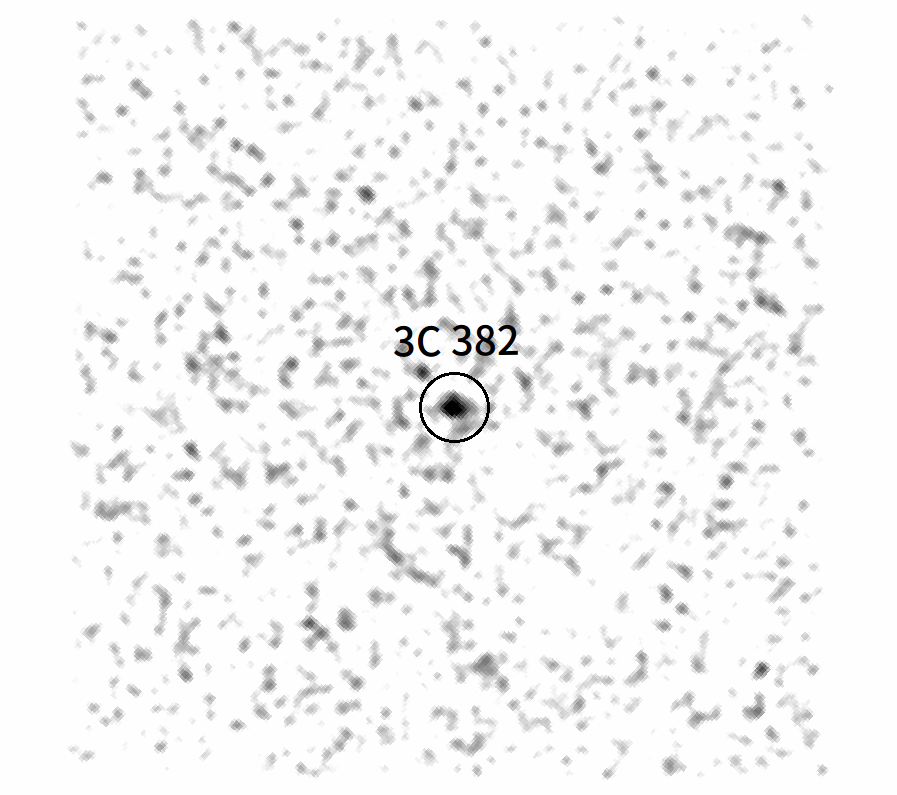}
\caption{The observed image of the quasar 3C 382, with 15 net source counts detected in the 604 seconds observation.}
\label{f3c382}
\end{figure}

\begin{figure}[htbp]
\centering
\includegraphics[width=1 \textwidth]{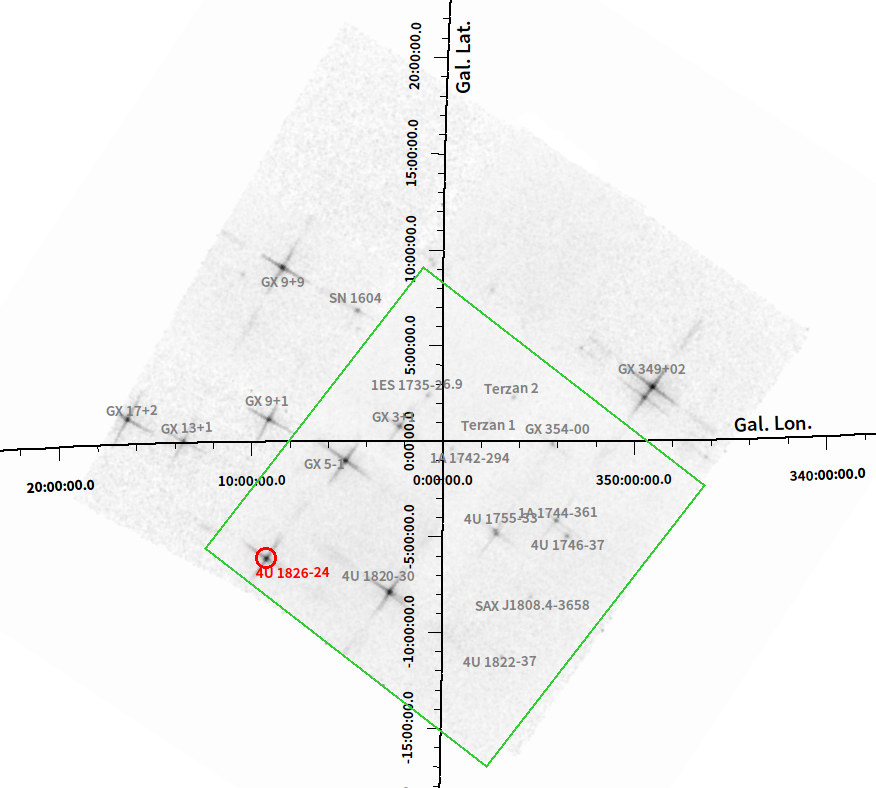}
\caption{A stacked image of 7 {\em LEIA} observations of the Galactic Center with an total exposure of 4652 seconds. All of sources detected  are known X-ray sources in this image. The blue box shows the Fov of {\em LEIA}.}
\label{gc7}
\end{figure}

\begin{figure}[htbp]
\centering
\includegraphics[width=0.8 \textwidth]{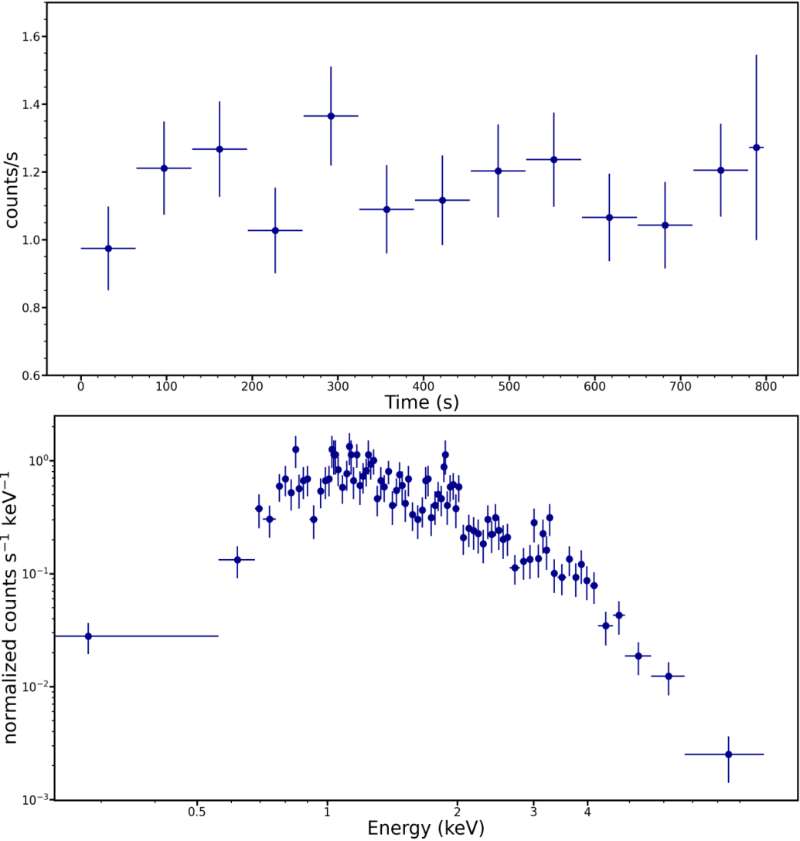}
\caption{The lightcurve and the spectrum of X-ray binary 4U1826-24 obtained in one pointing observation (marked by red circle in Figure \ref{gc7}).}
\label{gclightcurve}
\end{figure}

Furthermore, the fast readout and spectroscopic capability of the CMOS sensors make it possible to obtain the short-timescale X-ray light curves and energy spectra of bright sources during one pointed observation. 
One such example is shown in Figure \ref{gclightcurve}, where the X-ray lightcurve and spectrum of the X-ray binary 4U1826-24 can be obtained simultaneously in one of the observations in Figure \ref{gc7}.
These results demonstrate the great potential of monitoring X-ray sources and transients in the Galactic bulge region by a lobster-eye telescope such as {\em LEIA} with frequent visits. 

The imaging quality of {\em LEIA} has been examined with observations of Sco X-1 in the directions at the centers of one quadrant, giving PSF of 3.6 arcmin (FWHM). This is in excellent consistency with the PSF values measured in on-ground calibration. It is interesting that the X-ray image of diffuse X-ray sources can also be produced by lobster-eye optics. 
Example images of Sco X-1 and Cygnus Loop are presented in Figure 4 in \cite{Zhang2022}.

The on-board triggering system and the algorithm were also tested during the PV phase.  
Their performance is still under evaluation and the result will be presented elsewhere in the future (Liu et al. in preparation).

A observation of 3C 382 was conducted. The observed image is shown in Figure \ref{f3c382}. There are net 15 X-ray photons with a significance of 6.3 in a diameter of 8 arcmin region in a 604 second observation. The flux of 3C 382 is about 3.4 to 4.2 $\times 10^{-11} \rm{\ erg\ s^{-1}\ cm^{-2}}$ in 2 - 10 keV \citep{2018MNRAS.478.2663U}. In spite of the uncertainty introduced by the unknown X-ray spectral parameters of the quasar at the time of the observation, the result indicates that LEIA has a sensitivity consistent with that estimated from the simulations. A systematic analysis of the LEIA sensitivity is on-going with more observations on faint X-ray sources and will be presented elsewhere.

\subsection{In-orbit calibration}

The in-orbit calibration campaign has mainly been carried out during Oct 1 to Dec 17 2022 and some of the calibration observations are still on-going.
For the calibration of the effective areas and the source positioning offsets, the Crab nebula is the calibration source.
The Crab nebula has been observed a large number of times, in each of which the telescope pointing is adjusted in such a way that the Crab image falls onto a given position (direction) on a $22 \times 22$ grid within the entire FoV. In this way the effective areas and positioning offsets can be calibrated for a large number of directions of the FoV.   
For other directions not sampled, interpolation is to be performed.
For the calibration of the energy resolution and gain of the detectors, Cas A is the primary source, and the calibration is still on-going. 
The data of the in-orbit calibration observations of {\em LEIA} are still under extensive analysis and the results will be presented in a separate paper.  

\subsection{Current status}
Since the start of its operation in 2022 August, {\em LEIA} has been operating in orbit for about 6 months so far. Preliminary analysis shows that there is no significant degradation of its performance over the 6 months of operation.  

From 2022 October, {\em LEIA} has started test scientific observations, in addition to the on-going calibration observations of the Crab nebula and other tests. 
There are several observation programs being undertaken, including monitoring the Galactic plane, the Large and Small Magellanic Clouds (LMC, SMC), and surveys of selected high Galactic latitude sky regions.  
However, the scientific productivity of {\em LEIA} could be severely limited by a number of factors. The first is the relatively short observable time intervals (10 -- 15 minutes) largely owing to high particle background in the unfavorable orbit of SATech-01. Besides, several of the daily orbits will be unavailable for observations due to data telemetry of the satellite and other constraints. Secondly, the latency of receiving data from data taking is about one day at the moment,  which is too long for some short-lived transients to be identified and their alerts to be triggered.  
A lack of on-board telescope also hampers quick follow-ups once a fast transient is detected.  

\begin{figure}[htbp]
\centering
\includegraphics[width=1 \textwidth]{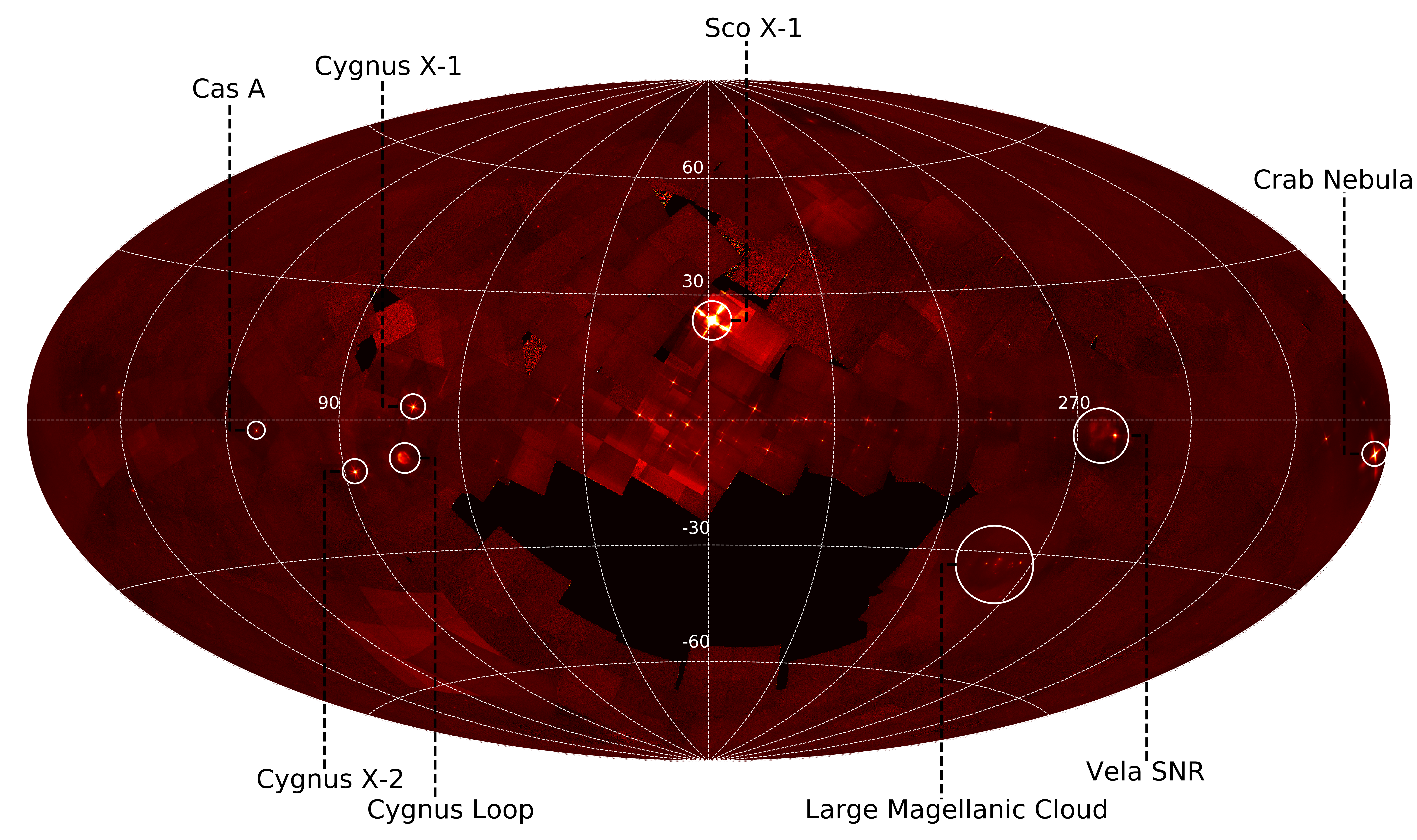}
\caption{The sky coverage of {\em LEIA} observations as of April 2023 (in Galactic coordinates).}
\label{map}
\end{figure}

Albeit these drawbacks, {\em LEIA} has being conducting scientific observations since 2022 October and has been accumulating an increasing amount of data set.
As of the time of writing, {\em LEIA} has made a total of 1542 pointed observations, covering a total sky region of 36000 square degrees. The average exposure time of one pointed observation is 868 seconds and the total observing time amounts 1.34 Ms.
An extensive analyses of this data set is to be carried out, especially when the in-orbit calibration is completed and the calibration data are fully analyzed.  
Here we only highlight several preliminary results.
Thanks to the large FoV,  {\em LEIA} has surveyed the Galactic plane for 10 times, covering a region of $l = 0 - 270$ and $l = 345 - 360$ degrees and $b = -15 - 15$ degrees.
The FoV of {\em LEIA} can fully cover LMC in one snapshot, and it also carried out monitoring observations of LMC for 90 times in the past several months. On 2023 January 20, an outburst of LMC X-4 with an increase of its X-ray flux by a factor of 30 was detected by {\em LEIA} \citep{2023ATel15871....1L}.
{\em LEIA} is also performing a large-scale sky survey at high Galactic latitude regions, aiming at detecting bright extra-galactic variables and transients.
So far, more than half of the entire sky has been covered, mostly with one pointing lasting for 10 -- 15 minutes.
Figure \ref{map} shows the sky coverage by {\em LEIA} in Galactic coordinates until April 2023.
It is expected with the motion of the Sun in the sky the rest of the sky regions will be covered in the next several months. 

{\em LEIA} has also detected a number of transient events or candidates. As an example, 
during one of the calibration observations of Crab, a bright transient source LXT221107A, the very first one for {\em LEIA}, was detected in the same FoV \citep{2022ATel15748....1L}. Follow-up observations with Swift/XRT also detected the source and enabled precise localization and identification with a nearby star HD 251108 at a distance of $497 \pm 6$ pc \citep{2022arXiv220800211G}. The peak luminosity observed with {\em LEIA} is $1.5^{+0.4}_{-0.2} \times 10^{34}\, \rm{erg\ s^{-1}}$, making it one of the brightest stellar X-ray flares. A detailed analysis of this flaring event is underway. 
In addition, several other fast X-ray transients or candidates were also detected \citep{2022ATel15834....1C}. 
More transient and variables are expected to be detected in future observations.

 When the on-board triggering system is turned on, it will  search for transients automatically. Once a transient is found, the alert message of the source will be transmitted to the ground by the Beidou navigation system in real time. The information about the new transient will be examined at the Einstein Probe Science Center based at NAOC. For genuine transient sources of potentially high scientific interests, alerts will be publicly released to the wider astronomical community to call for multiwavelength follow-up observations.

\section{Conclusions}

As a pathfinder of the Wide-field X-ray Telescope of the Einstein Probe mission of the CAS, {\em LEIA} is a lobster-eye X-ray imager, which was launched on 2022 July 27 onboard the CAS's SATech-01 satellite. 
The primary goals of {\em LEIA} are to verify the in-orbit performance of the state-of-the-art MPO and X-ray CMOS devices, the two novel technologies that are key to the EP mission.
With a FoV of 346 square degrees, it is the first wide-FoV X-ray focusing telescope operating in orbit.
Previous on-ground calibrations demonstrated that the imaging PSF (4 -- 8 arcmin) and effective areas (2 -- 3 $\rm{cm^2}$ for the central spot at 1.25 keV) are largely uniform with only mild variations across the entire FoV except at the edges. 
Thus it is a truely wide-FoV X-ray focusing telescope with only little vignetting effects. 
{\em LEIA} has a nominal soft X-ray bandpass of 0.5 -- 4 keV.

Since its operation in orbit, extensive testing and calibration observations have been performed in the following months. So far, the instrument works mostly normally as far as its main functions are concerned, focusing imaging, spectroscopy and timing. 
Preliminary results derived from observations taken in this period suggest that both the MPO and CMOS sensors adopted have been working well.
The measured FoV, PSF and effective areas are well consistent with those measured in previous on-ground calibrations.
The typical detecting sensitivity of {\em LEIA} is $2 - 3 \times 10^{-11}$ $\rm{erg\ s^{-1}\ cm^{-2}}$ for 1000 second exposure, depending on the background level.
This value is roughly consistent with the simulated result based on on-ground calibration data.
In-orbit calibrations have been carried out and the data analysis is to be completed in the near future. 
Preliminary results show that no significant degradation of the performance has been observed in the past 5-month observations. 

These results demonstrate that the lobster-eye MPO, CMOS sensors, and the entire telescope are proved to work, and the goals of {\em LEIA} have largely been achieved. 
This is also the first time, to our knowledge, that CMOS sensors are applied to X-ray astronomical detection in space. The {\em LEIA} experiment will pave the way for the development of the present and future X-ray missions using lobster-eye and CMOS technologies.

The available observing intervals are relatively short, typically about 10 -- 15 minutes for each usable orbit, while the latency of data telemetry is some what long. In spite of this, the wide FoV combined with the moderate angular resolution and sensitivity enabled by lobster-eye MPO make {\em LEIA} a promising soft X-ray wide-field monitor. This has been evidenced by the detection of several transients and outbursts by {\em LEIA} in the past several month-operation. {\em LEIA} is expected to carry out monitoring observations of the Galactic plane, LMC/SMC, and selected sky regions at high Galactic latitudes in the following 1.5 years or so that is expected from its designed lifetime.

\begin{acknowledgements}
This work is supported by the
Einstein Probe project, a mission in the Strategic Priority
Program on Space Science of CAS (grant Nos. XDA15310000,
XDA15052100). We acknowledge contribution
from Leicester and MPE teams (funding by ESA) for calibration of the mirror assembly and tests of part of the devices. The work performed at MPE’s PANTER X-ray test facility has in part
been supported by the European Union’s Horizon 2020 Program under the AHEAD2020 project (grant No. 871158).

\end{acknowledgements}
  
\bibliographystyle{raa}
\bibliography{LEIA_Design_RAA}

\label{lastpage}

\end{document}